\documentclass{jpp}
\usepackage{epstopdf, epsfig}

\usepackage[utf8]{inputenc}
\usepackage[T1]{fontenc}
\usepackage{amsmath} 
\usepackage{graphicx} 
\usepackage{natbib}
\usepackage{hyperref}
\usepackage{xcolor}
\usepackage{ulem}

\newcommand{\Bv}{{\bf B}}
\newcommand{\Ev}{{\bf E}}
\newcommand{\fv}{{\bf f}}
\newcommand{\uv}{{\bf u}}

\newcommand{\vv}{{\bf v}}

\newcommand{\Hv}{{\bf H}}

\newcommand{\rv}{{\bf r}}
\newcommand{\Yv}{{\bf Y}}
\newcommand{\jv}{{\bf j}}

\shorttitle{Energy transfer in plasma turbulence}
\shortauthor{C. L. V\'asconez et al.}

\title{Local and global properties of energy transfer in models of plasma turbulence}

\author{Christian L. V\'asconez\aff{1}
  \corresp{\email{christian.vasconez@epn.edu.ec}},
  D. Perrone\aff{2}, R. Marino\aff{3}, D. Laveder\aff{4}, F. Valentini\aff{5}, S. Servidio\aff{5}, P. Mininni\aff{6} \and L. Sorriso-Valvo\aff{7}}

\affiliation{\aff{1}Departamento de F\'isica, Escuela Polit\'ecnica Nacional, Ladr\'on de Guevara E11-253, 170525 Quito, Ecuador
\aff{2}ASI -- Italian Space Agency, via del Politecnico snc, 00133 Rome, Italy
\aff{3}Laboratoire de M\'ecanique des Fluides et d’Acoustique, CNRS, \'Ecole Centrale de Lyon, Universit\'e Claude Bernard Lyon 1, INSA de Lyon, Écully F-69134, France
\aff{4}Universit\'e C\^ote d’Azur, CNRS, Observatoire de la C\^ote d’Azur, Laboratoire J. L. Lagrange, Boulevard de l’Observatoire, CS 34229, 06304 Nice Cedex 4, France
\aff{5}Dipartimento di Fisica, Universit\`a della Calabria, I-87036, Rende (CS), Italy
\aff{6}Departamento de F\'isica, Universidad de Buenos Aires and IFIBA, CONICET, 1428 Buenos Aires, Argentina
\aff{7}Istituto per la Scienza e Tecnologia dei Plasmi (ISTP), Consiglio Nazionale delle Ricerche, Via Amendola 122/D, 70126 Bari, Italy}

\begin{document}

\maketitle

\begin{abstract}
The nature of the turbulent energy transfer rate is studied using direct numerical simulations of weakly collisional space plasmas. This is done comparing results obtained from hybrid Vlasov-Maxwell simulations of colissionless plasmas, Hall-magnetohydrodynamics, and Landau fluid models reproducing low-frequency kinetic effects, such as the Landau damping. In this partially developed turbulent scenario, estimates of the local and global scaling properties of different energy channels are obtained using a proxy of the local energy transfer (LET). This approach provides information on the structure of energy fluxes, under the assumption that the turbulent cascade transfers most of the energy that is then dissipated at small scales by various kinetic processes in this kind of plasmas.
\end{abstract}

\section{Introduction}

Space plasmas are a unique laboratory to study the transfer of energy in highly turbulent media \citep{bruno2016turbulence}. In particular, the Solar Wind near the Earth has been continuously probed by space missions \citep{tu1995mhd}. The {\it in situ} measurements of this quasi-collisionless, highly variable, and structured plasma show a medium in a state of fully-developed turbulence. Under these conditions, turbulence is primarily originated at the Sun, and transported at high speed away from the source \citep{goldstein1996ulysses}. This turbulence is then dissipated through magnetohydrodynamic (MHD)- to kinetic-scales, where processes as plasma waves excitation (e.g. \citep{pezzi2013eulerian}), temperature anisotropy (e.g., \citep{servidio2012local,perrone2014generation}), plasma heating (e.g., \citep{smith2001heating,perrone2014analysis,vaivads2016turbulence}), particle energization (e.g., \citep{gibelli2010ion}), entropy cascade (e.g., \citep{cerri2018dual,yang2017energy}), and enstrophy cascade (e.g., \citep{servidio2008depression}) are activated. 
Consequently, this energy cascade produces finer structures in the particles velocity distribution function (VDF) \citep{marsch2004temperature,valentini2008cross,vasconez2014vlasov}.

The MHD approximation describes space plasmas phenomenology at large enough scales \citep{servidio2008depression}. In this approach, a Kolmogorov-like behaviour is highly supported by observations of velocity and magnetic fluctuations showing power-law spectra and intermittency \citep{carbone2004intermittency,greco2009statistical}. At the MHD scales, the Solar Wind has shown scale-dependent non-Gaussian statistics \citep{sorriso1999intermittency}, with non-Gaussian fluctuations observed as well in the large scales \citep{marino2012} as it happens also in anisotropic fluids with waves \citep{feraco2018}. Then, its turbulent energy is dissipated in an efficient way as a result of the inhomogeneous energy transfer, mostly occurring at small-scale vorticity filaments, rotational (or tangential) discontinuities, and current sheets, among other structures \citep{zimbardo2010magnetic}.          
At scales close the proton inertial length and/or to the proton skin depth, pure MHD models are no longer valid \citep{matthaeus2008comment}. Kinetic processes, led by field-particle interactions, have to be considered. Observations at $1$AU show non-Maxwelliam VDFs of ions and electrons in the zone where a low collisions rate is measured \citep{leamon1998observational}. Complementary to the observations, Vlasov-Maxwell numerical simulations have confirmed that particles energization is taking place near the most intense small-scale structures \citep{servidio2015kinetic}. The processes responsible for such energization are not yet understood. However, mechanisms as magnetic reconnection (e.g., \citep{servidio2009magnetic}), plasma instabilities (e.g.,  \citep{primavera2019parametric,settino2020proton}), wave-particle interactions (e.g. \citep{sorriso2019turbulence,chen2019evidence} and increase of collisions (e.g., \citep{pezzi2016alfvenic,pezzi2017turbulence}) have been pointed out as good candidates. 
Then, direct numerical simulations have reveled to be quite useful to understanding the physics of the plasmas under conditions of the near-Earth Solar Wind. 

In this work, we study the properties of the turbulent energy transfer in plasmas in the direct numerical simulations (DNS) framework previously investigated in \citet{perrone2018}. In their work, Hall-magnetohydrodynamic (HMHD), Landau Fluid (LF), and hybrid Vlasov-Maxwell (HVM) bidimensional simulations in turbulent regimes were ran under collisionless-plasma conditions, considering an out-of-plane ambient magnetic field. Magnetic diffusivity was carefully introduced in the fluid models. The fields obtained from these simulations allow us to explore different scales linked to their respective range of validity. 

This paper is organized as follows. In Section \ref{dns}, we briefly present the three numerical models, namely HMHD, LF and HVM, which describe --under the same 2D configuration-- a collisionless plasma, in typical conditions of the Solar-Wind plasma, in a quasi-developed turbulent state. Our analysis of local and global energy transfer of this state is shown in Section \ref{let} and Section \ref{global}, respectively. We summarize and conclude in Section \ref{conclusion}.  


\section{Numerical models}
\label{dns}

The dynamical behavior of a plasma strongly depends on its frequencies. Here, we provide a very brief description of the three models used in the present work, which can properly focus on different ranges of frequency. At the lowest frequencies, where ions and electrons are locked together, and the plasma behaves like an electrically conducting fluid, a MHD model is good enough to describing the system. In this study, we use a HMHD simulation, which retains dispersive effects that become relevant at scales comparable to the proton skin depth. However, when collisions can be neglected, pressure anisotropy can develop in the plasma. In this case, and at somewhat higher frequency than the previous regime, a two-fluid description is needed, since electrons and ions can move relatively to each other. Therefore, we use a LF model which is able to take into account both pressure anisotropy and low-frequency kinetic effects, such as Landau damping. At higher frequencies, a kinetic model is required. We consider the HVM approach which describes the evolution of the proton distribution function, while electrons are treated as a fluid. All the three models include the electron inertial effects with a protons-to-electron mass ratio $m_p /m_e =100$.

The HMHD and LF codes use a 2D Fourier pseudo-spectral method to compute the spatial derivatives, while the time advance is performed with a third-order Runge-Kutta scheme. In these fluid codes, aliasing errors in the evaluation of nonlinear terms are treated by a $2/3$ truncation in the spectral space. In the HVM code, time advance is based on the time-splitting scheme \citep{cheng1976integration}, combined with the Current Advance Method --CAM \citep{matthews1994current}. In the Vlasov equation, advection in space and velocity are solved through an explicit upwind scheme \citep{van1977towards}. The HVM code uses FFT's operations only when computing their fields, and truncation is not needed since possible dealiasing errors are damped by the intrinsic dissipation of the numerical scheme.

The vector fields will evolve in a double periodic domain $D = [L,L]$, in the $(x,y)$ plane. $L = 2\pi \times 25 d_p$ is the length of each direction $x$ and $y$. This domain is discretized with $N_x = N_y = 1024$ grid points. Additionally, the HVM simulation is run in a 3D velocity domain $- 5 v_{th,p} \leq \vv \leq 5 v_{th,p}$, discretized with $51^3$ points, and where $v_{th,p}$ is the thermal velocity of the protons. At $t=0$, the equilibrium was set by a homogeneous plasma embedded in an uniform background magnetic field, $\Bv_0 = (0,0,B_0)$, oriented in the $z$-direction. The initial perturbation of this equilibrium is performed by velocity (and magnetic) field fluctuations with random phases in the interval $2 \leq m \leq 6$, for the wavenumbers $k = 2 \pi m /L$. Although the amplitude of the magnetic perturbations are $b / B_0 \approx 0.3$, no density fluctuations neither parallel variances are imposed. In fluid models, a magnetic diffusivity term with a coefficient $\eta = 2 \times 10^{-2}$ has been added. 
\subsection{Hall-magnetohydrodynamic}
\label{hmhd}

The HMHD model includes the electron inertia contribution. The adimensional set of equations is
\begin{equation}
  \frac{\partial n}{\partial t} + \nabla \cdot (n \uv) = 0;
\end{equation}
\begin{equation}
  \frac{\partial \uv}{\partial t} + (\uv \cdot \nabla) \uv = - \frac{\beta}{2 n} \nabla P + \frac{1}{n} \left[ (\nabla \times \Bv) \times \Bv \right];    
\end{equation}
\begin{equation}
  \frac{\partial \Bv}{\partial t} = -\nabla \times \Ev + \eta \nabla^2 \Bv;
 \label{eq:faraday}    
\end{equation}
\begin{equation}
  (1 + \alpha \nabla^2)\Ev + \uv \times \Bv -\frac{1}{n} \left[ (\nabla \times \Bv) \times \Bv \right] = 0;\\
 \label{eq:ohm}
\end{equation}
\begin{equation}
  P = n^\gamma, 
\end{equation}    
\noindent
where $n$ is the plasma density, $\uv$ is the hydrodynamic velocity, $\Ev$ and $\Bv$ are the electric and magnetic field, respectively. $P$ is the isotropic total pressure, $\eta = 2 \times 10^{-2}$ is the magnetic diffusivity, $\gamma = 5/3$ is the adiabatic index, and $\alpha=1/100$ is an artificial electron to proton mass ratio. This adimensional set of equations are obtained normalizing $n$ to $n_0 m_p$, $\uv$ to the Alfv\'en speed $v_A$, $t$ to the inverse proton cyclotron frequency $\Omega_{cp}^{-1}$, and the length units to the proton skin depth $d_p = v_A / \Omega_{cp}$. For regularization purposes, and to reproduce the same-time behavior for the maximum value of the integrated parallel current density $\langle j_z^2\rangle$, bi-Laplacian hyperviscosity and hyperdiffusivity with coefficients equal to $5\times 10^{-4}$, have been added to the RHS of velocity and magnetic field equations.    
\subsection{Landau fluid}
\label{lf}

The LF system of dynamical equations is set for solving the evolution of the magnetic field, density, velocity, parallel and perpendicular pressures, and heat fluxes of the ions \citep{sulem2015}. The electric field is computed using a generalized Ohm’s law, which includes the Hall term, together with electron inertia with the same simplified form and the same artificial mass ratio as in the HMHD case. The electron pressure gradient, which is formally present in the Ohm’s law, is here neglected to closely match the conditions of the Vlasov simulations, in which electrons are cold. The fluid hierarchy is closed by the gyrotropic fourth-rank moments in terms of the above quantities, in a way consistent with the low-frequency linear kinetic theory. In this study we are not considering the ion finite Larmor radius, and only ion linear Landau damping is retained, in the approximation discussed in \citep{passot2014}. Together with the same hyperviscosity and hyperdiffusivity (as in HMHD equations), additional bi-Laplacian dissipative terms have been supplemented in the equations for the density and the pressures (with coefficient equal to $2.5 \times 10^{-3}$), and in the equations for the ion heat flux (with coefficient $10^{-4}$) in order to deal with the high level of compressibility in the simulation (the Mach number reaching values up to $0.4$). In fact, we note that compressibility of these fluid models will be important when comparing our results.

\subsection{Hybrid Vlasov-Maxwell}
\label{hvm}

The hybrid Vlasov-Maxwell (HVM) code \citep{valentini2007} solves the Vlasov equation for the protons distribution function $f = f (\rv, \vv, t)$, while the electrons response is taken into account through a generalized Ohm’s law for the electric field. The Vlasov equation, in normalized units, 
\begin{equation}
  \frac{\partial f}{\partial t} + \vv \cdot \frac{\partial f}{\partial \rv} + (\Ev + \uv_p \times \Bv) \cdot \frac{\partial f}{\partial \vv} = 0, 
  \label{eq:hvm}
\end{equation}
\noindent
is solved in a 2D-3V phase-space domain (two dimensions in physical space, and three dimensions in velocity space), coupling it with equations (\ref{eq:faraday}), and (\ref{eq:ohm}). Velocities $\vv$ and $\uv_p$ (protons bulk velocity) are scaled to $v_A$. Quasi-neutrality ($n = n_e = n_p$), and cold electrons, is considered. The time step is chosen in order to satisfy the Courant-Friedrichs-Lewy (CFL) condition for the numerical stability. Protons distribution function is initialized with a homogeneous-density Maxwellian function. In this procedure, displacement current is neglected in the Ampère law, making the assumption of low frequencies.  
%

\section{Local energy transfer analysis}
\label{let}

The analysis developed in this work will be performed in a period of maximal turbulence activity. As described in \citet{perrone2018}, this state is reached at $t^* = 60 \Omega_{cp}^{-1}$, for the same set of simulations considered here. They established $t^*$ after following $\langle j_z^2 \rangle$ in time, for HMHD, LF, and HVM descriptions, where the same $\beta$ was hold in order to ensure similar level of density fluctuations. 

\begin{figure}
    \centering
    \includegraphics[width=0.3\textwidth]{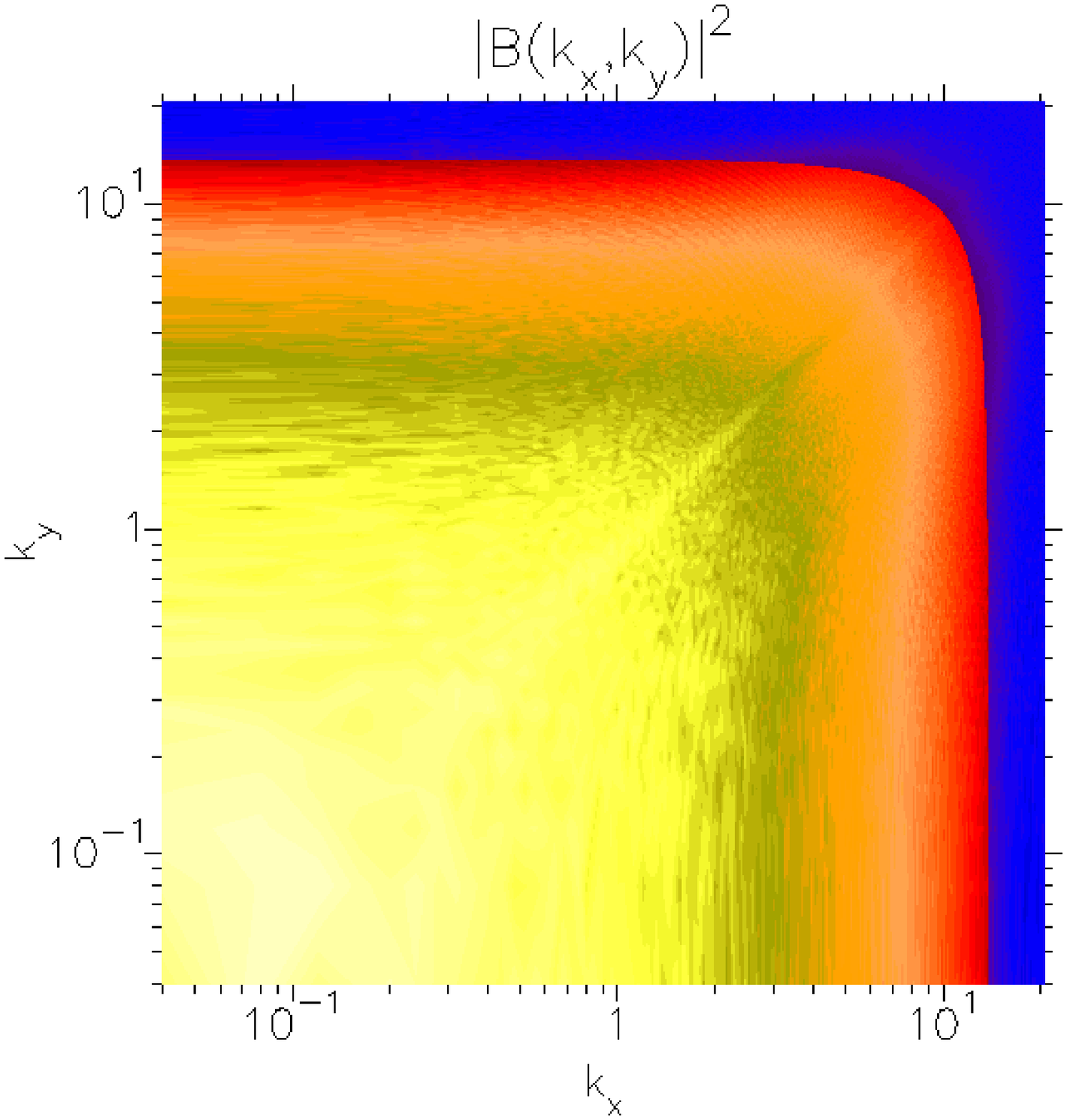}
    \includegraphics[width=0.3\textwidth]{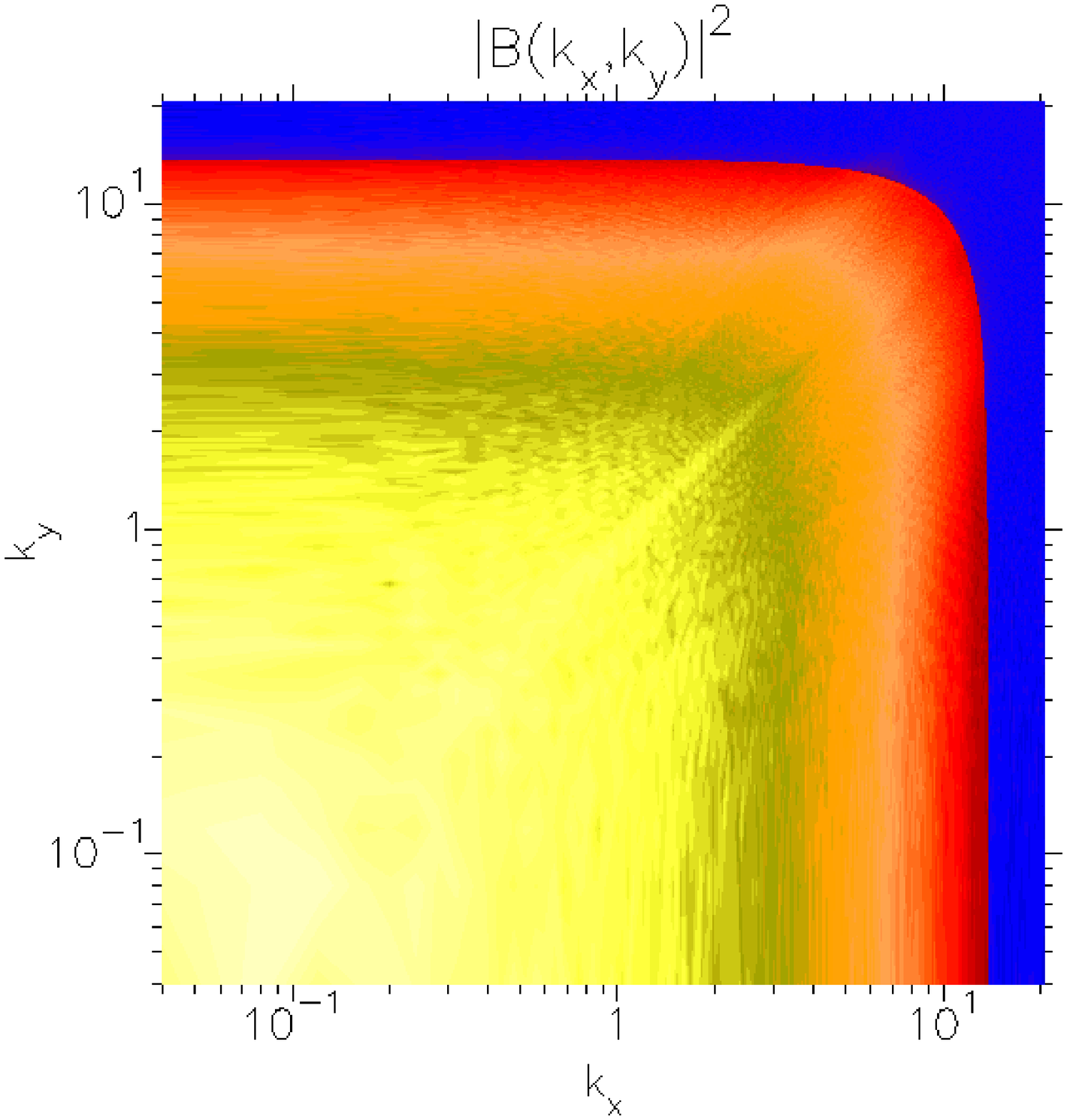}
    \includegraphics[width=0.3\textwidth]{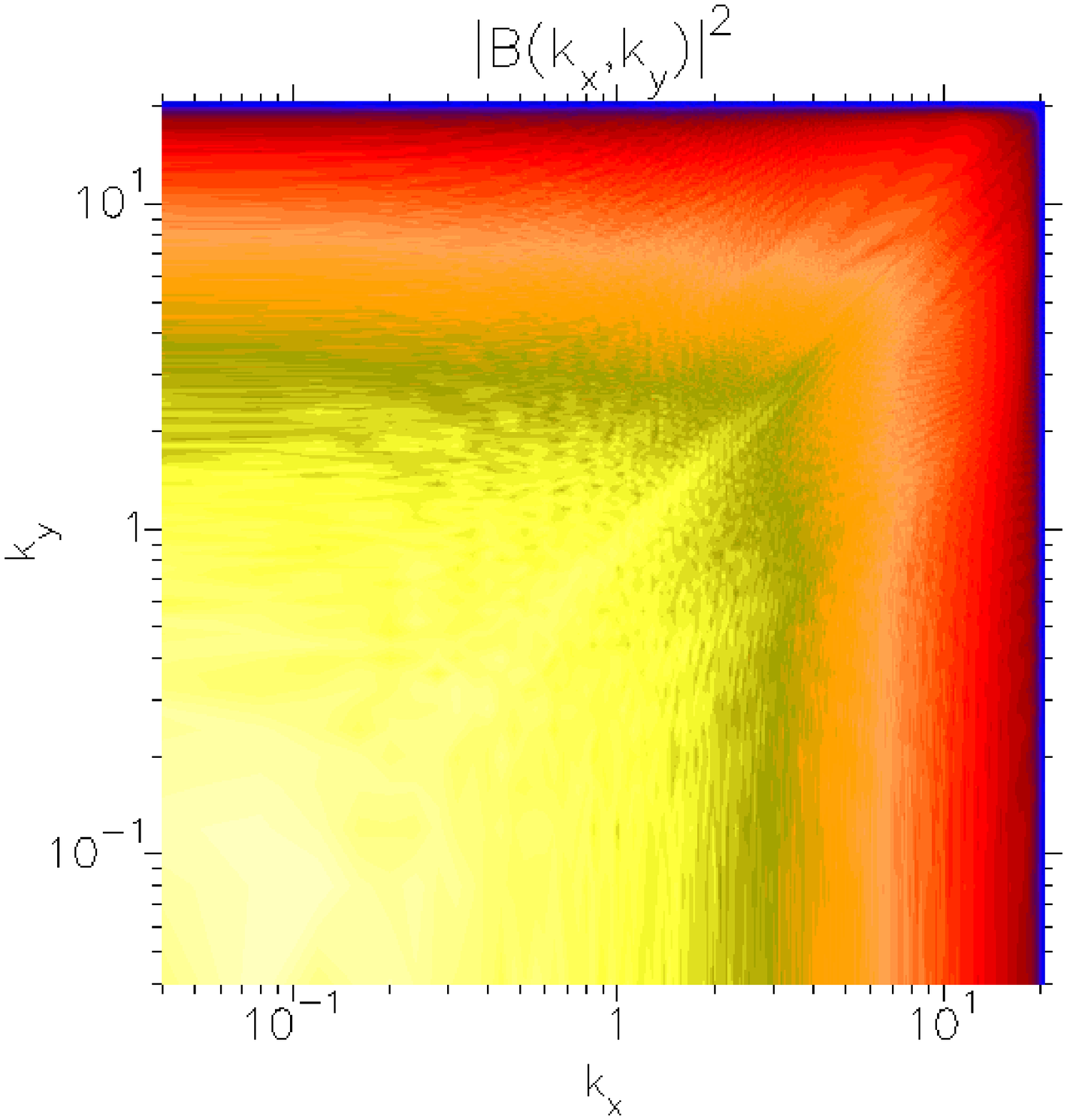}
    \includegraphics[width=0.3\textwidth]{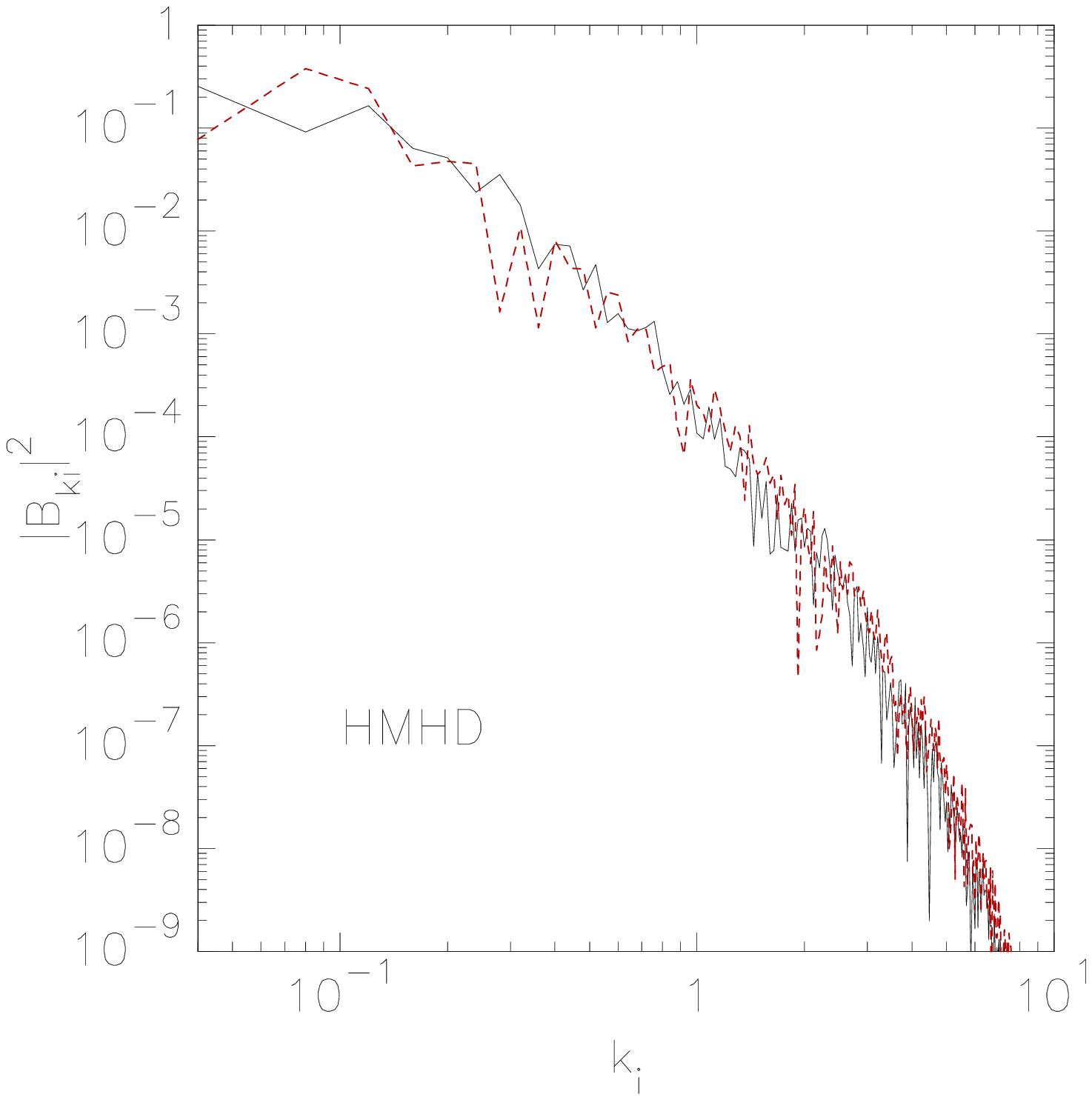} 
    \includegraphics[width=0.3\textwidth]{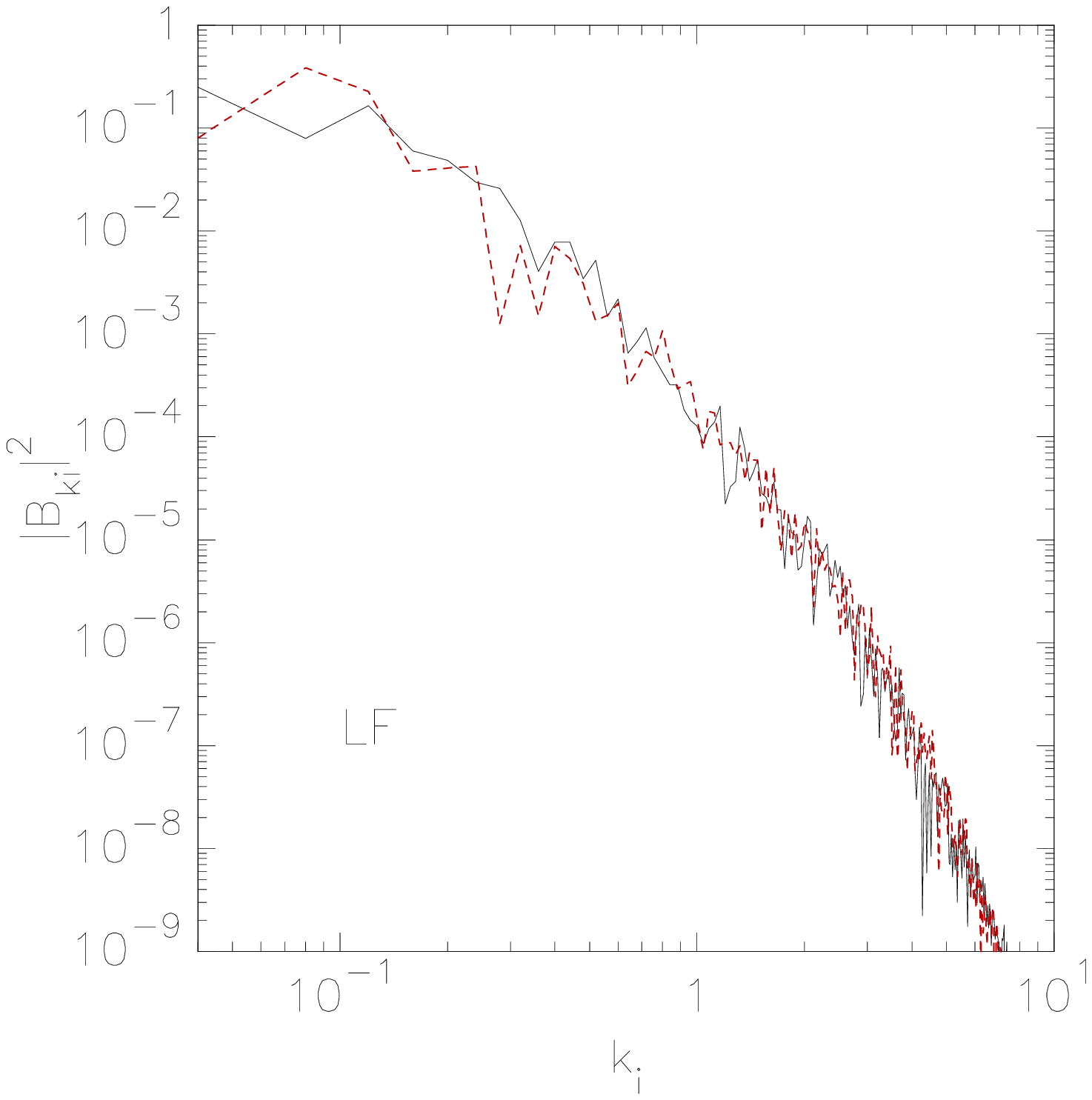} 
    \includegraphics[width=0.3\textwidth]{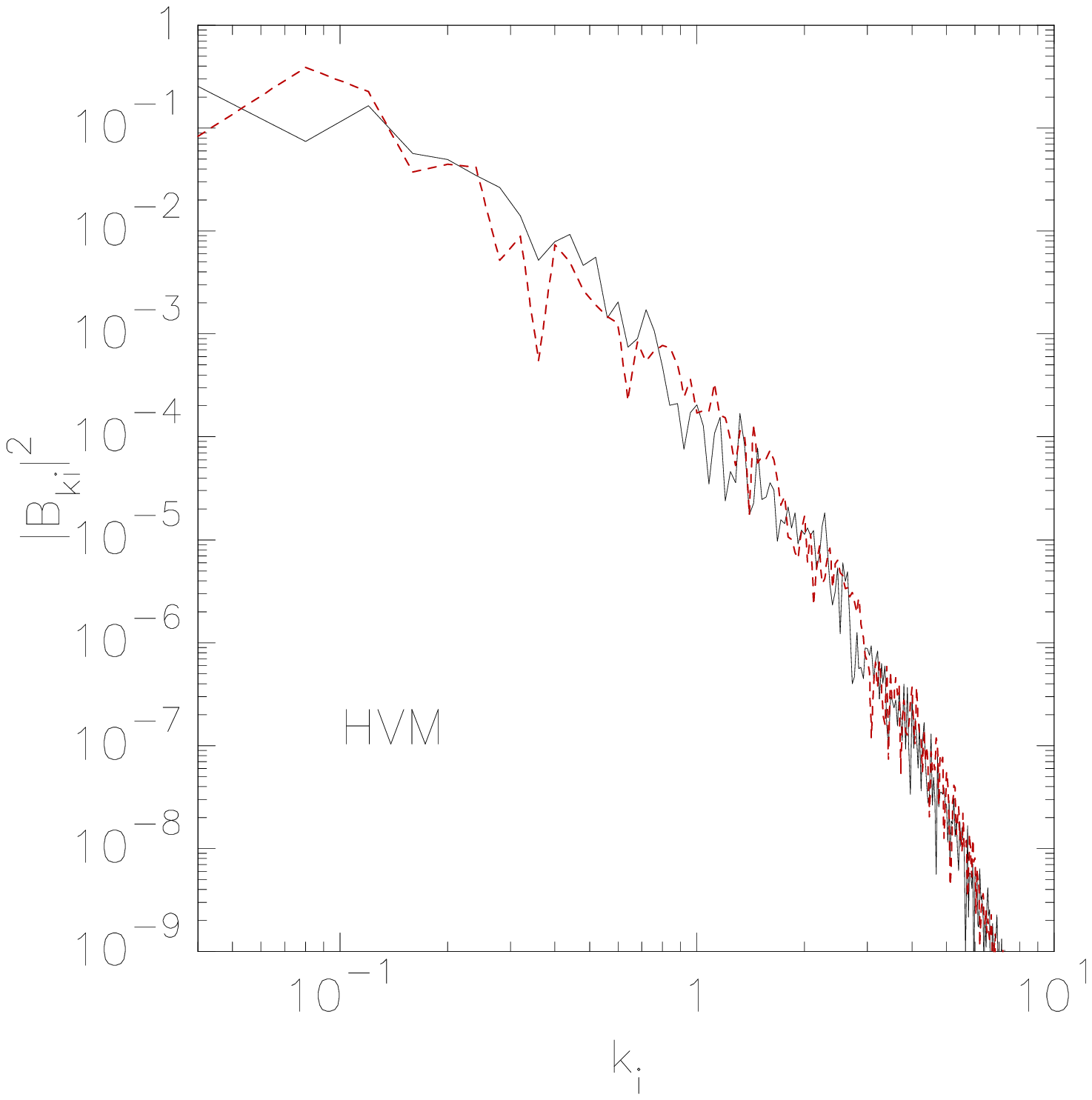}
    \includegraphics[width=0.3\textwidth]{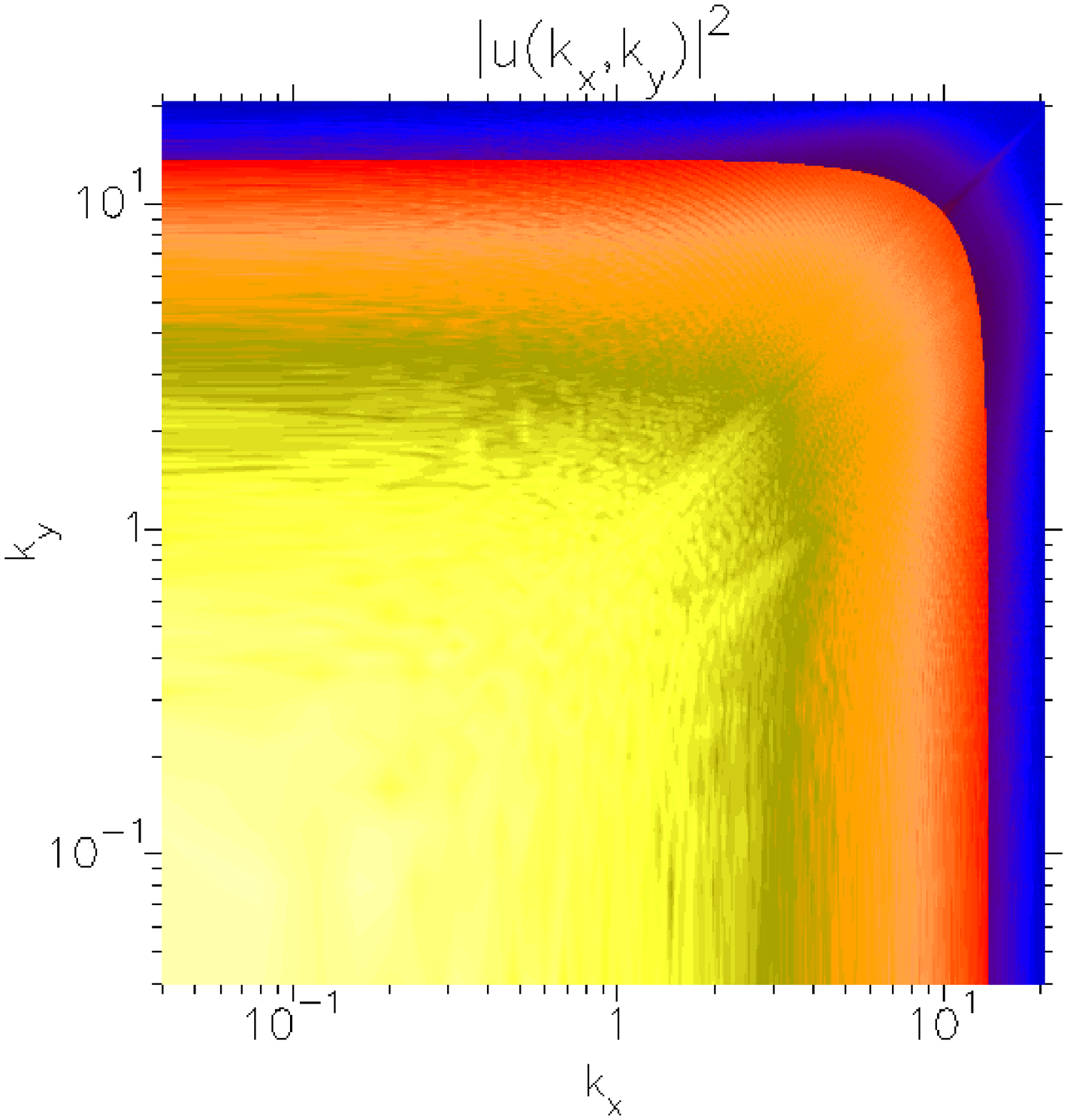}
    \includegraphics[width=0.3\textwidth]{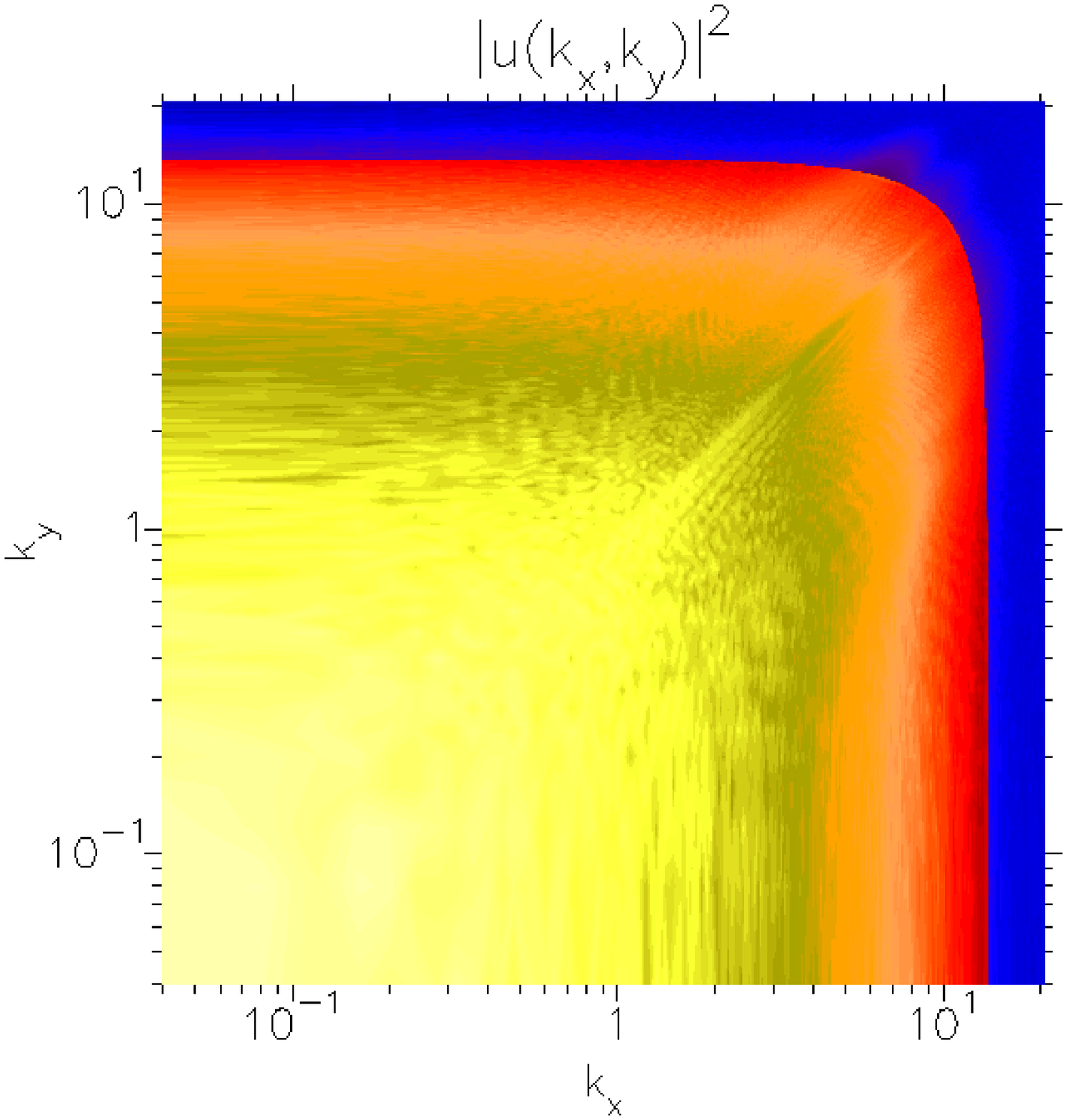}
    \includegraphics[width=0.3\textwidth]{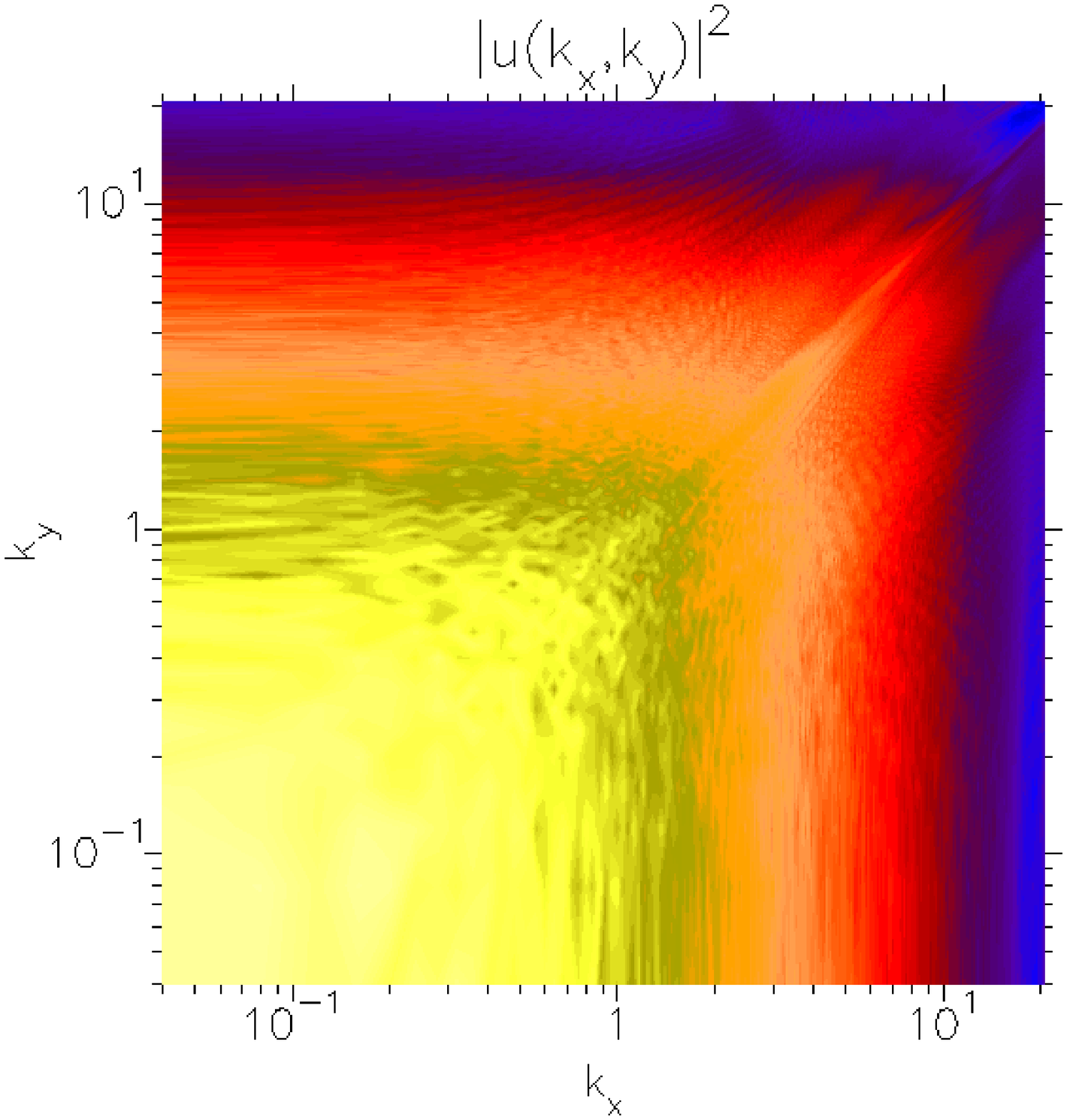}
    \includegraphics[width=0.3\textwidth]{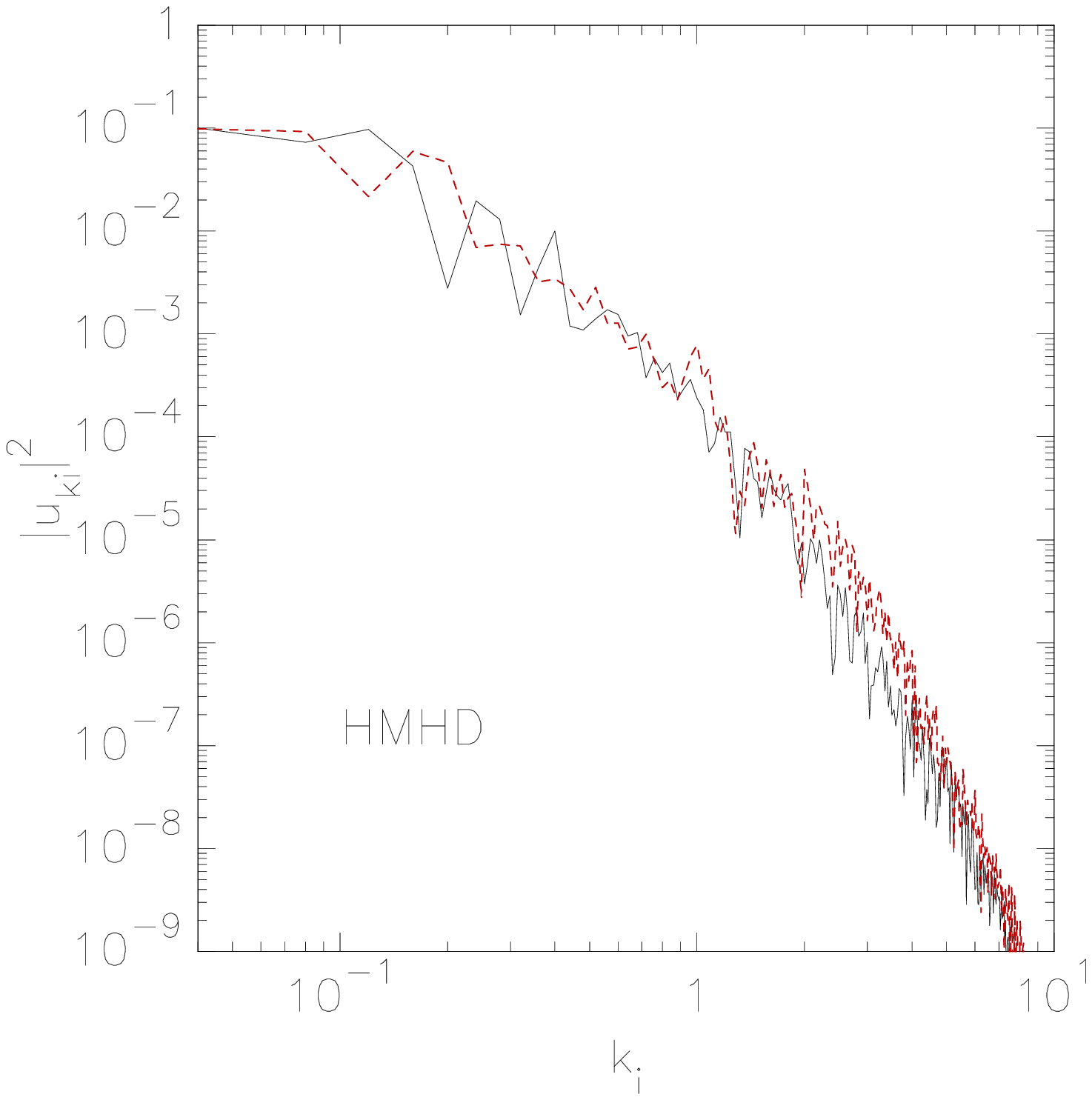} 
    \includegraphics[width=0.3\textwidth]{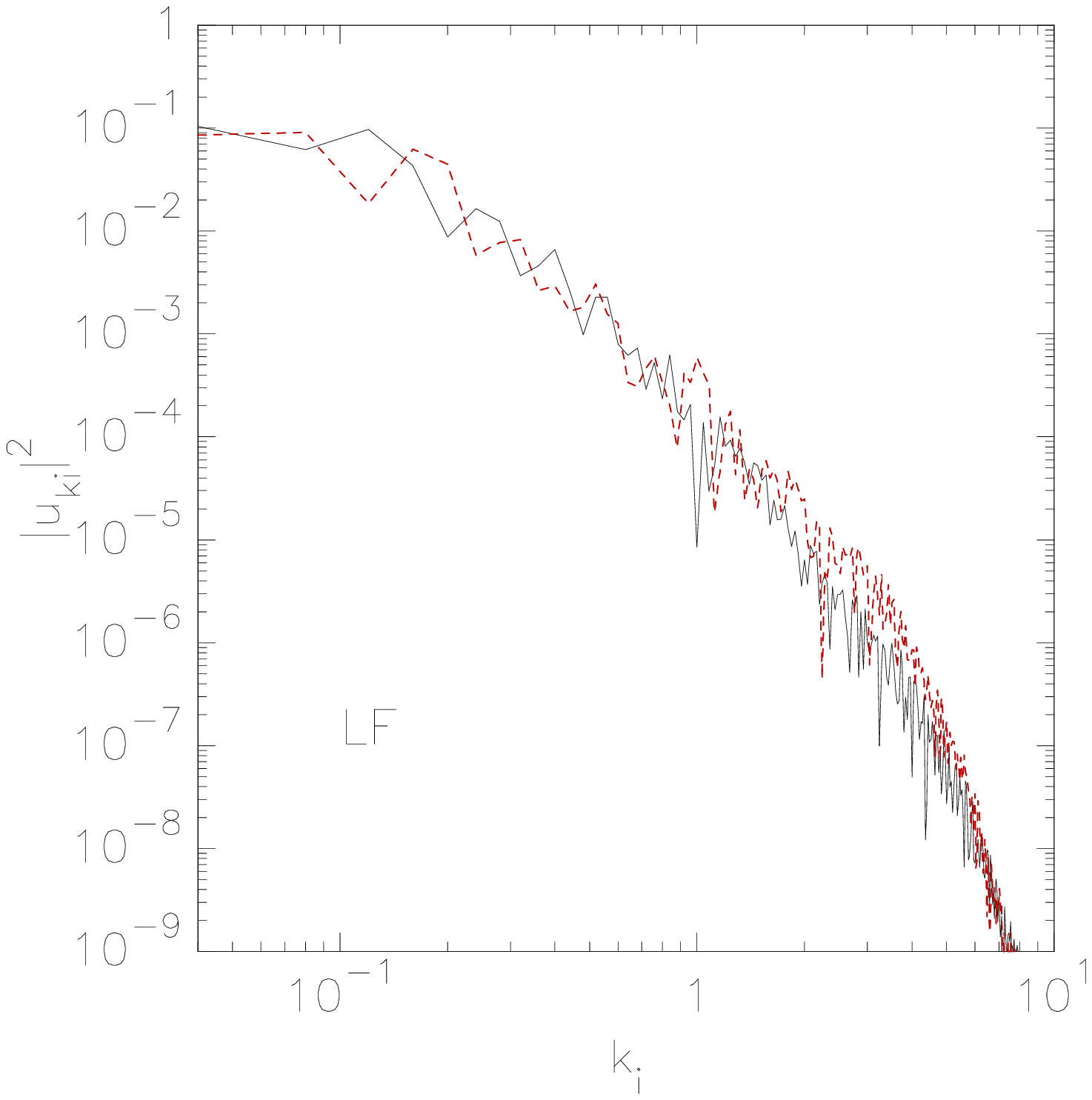} 
    \includegraphics[width=0.3\textwidth]{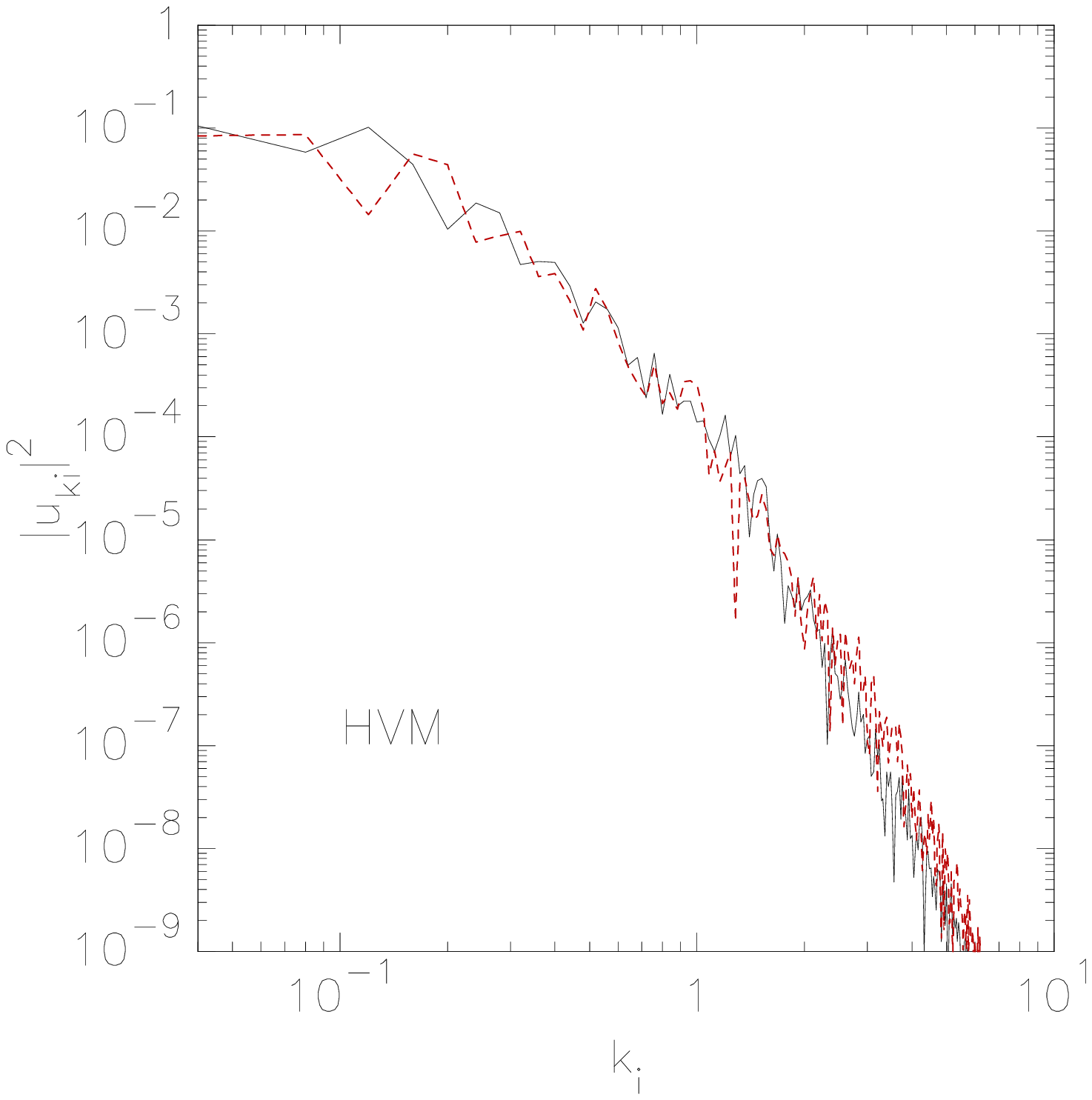} 
    \caption{Magnetic field (two-top rows) and velocity field (two-bottom rows) spectra. The integrated spectra are presented respect to $k_i$, $i = x$ (red-dashed line), $y$ (black-solid line).}
    \label{fig:spectra}
\end{figure}

Figure \ref{fig:spectra} shows the spectra of $\Bv$ and $\uv$, for the HMHD (left column), LF (middle column), and HVM (right column) simulations, respectively. Magnetic field spectra (bidimensional $|B(k_x,k_y)|^2$, and reduced integrated $|B_{ki}|^2$) are presented in the two-top rows, while the velocity field spectra (bidimensional $|u(k_x,k_y)|^2$, and reduced integrated $|u_{ki}|^2$) can be seen in the two-bottom rows. The integrated spectra are plotted respect to $k_i = k_x$ with a red-dashed line, and respect to $k_i = k_y$ with a black-solid line. Bidimensional spectra maps of HMHD, and LF simulations shows the $2/3$ truncation, which was not performed in the HVM simulation.    

Considering that, in the small-scales limit, HMHD physics was observed to match hybrid-kinetic models results (e.g., \citep{vasconez2015,pucci2016}), \citet{hellinger2018} presented a generalization of the von K\'arm\'an-Howarth equation for incompressible hydrodynamic turbulence, in the framework of incompressible HMHD equations. This generalization was revisited by \citet{ferrand2019}, who also considered homogeneous turbulence, and worked with two-point correlation tensors depending only on the relative displacement ${\bf \ell}$, and not on the absolute positions, as $\delta \fv = \fv (\rv + {\bf \ell}) - \fv (\rv)$. In this context, the mean rate of total energy injection $\epsilon$ is expressed as the combination of the third-order structure functions corresponding to the MHD turbulent cascade flux \citep{carbone2009,verdini2015}, and to Hall corrections (e.g., \citep{galtier2008}).
\begin{equation}
   - 4 \epsilon = \nabla \cdot \langle (\delta \uv \cdot \delta \uv) \delta \uv + (\delta \Bv \cdot \delta \Bv) \delta \uv - 2 (\delta \uv \cdot \delta \Bv) \delta \Bv - \frac{d_p}{2} (\delta \Bv \cdot \delta \Bv) \delta \jv + d_p (\delta \Bv \cdot \delta \jv) \delta \Bv \rangle.
   \label{eq:ferrand}
\end{equation}
In fact, we can identify the single contribution of each component that corresponds to the Yaglom-,
\begin{eqnarray}
 \Yv_1 (x,y) &=& (\delta \uv \cdot \delta \uv) \delta \uv \label{def:y1};\\
 \Yv_2 (x,y) &=& (\delta \Bv \cdot \delta \Bv) \delta \uv \label{def:y2};\\
 \Yv_3 (x,y) &=& -2 (\delta \uv \cdot \delta \Bv) \delta \Bv \label{def:y3},
\end{eqnarray}
\noindent
and to the Hall-effect contributions, 
\begin{eqnarray}
 \Hv_1 (x,y) &=&  d_p (\delta \Bv \cdot \delta \jv) \delta \Bv; \label{def:h1}\\
 \Hv_2 (x,y) &=& - \frac{d_p}{2} (\delta \Bv \cdot \delta \Bv) \delta \jv. \label{def:h2}
\end{eqnarray}                         
\noindent
In this way, equation \ref{eq:ferrand} would be written as $-4 \epsilon = \nabla \cdot \langle  \Yv_1 + \Yv_2 + \Yv_3 + \Hv_1 + \Hv_2 \rangle$. \\

On the other hand, a heuristic proxy has been recently introduced. It focuses on the local turbulent energy transfer rate (LET) towards the smallest resolved scale. The proxy was constructed in order to extend the Yaglom law to MHD turbulence, and matches the latter law when small density fluctuations (at the scale $\ell$) can be neglected~\citep{carbone2009scaling}. It is estimated, for 2D incompressible turbulence, through the combined third-order fluctuations of velocity, magnetic field, and current density~\citep{sorriso2018local}, and neglecting unity-order multiplicative factors, as 
%
%
\begin{equation}
   - 2 \ell \epsilon_\ell \equiv (\delta \uv \cdot \delta \uv) \delta u_\ell + (\delta \Bv \cdot \delta \Bv) \delta u_\ell - 2 (\delta \uv \cdot \delta \Bv) \delta B_\ell - \frac{d_p}{2} (\delta \Bv \cdot \delta \Bv) \delta j_\ell + d_p (\delta \Bv \cdot \delta \jv) \delta B_\ell.
   \label{eq:let}
\end{equation}
Consistently with our notation, the latter equation could be rewritten as
\begin{equation}
 -2 \ell \epsilon_\ell = \epsilon_Y + \epsilon_H,   
\end{equation}
\noindent
where $\epsilon_Y = \epsilon_{Y_1} + \epsilon_{Y_2} + \epsilon_{Y_3}$, and $\epsilon_H = \epsilon_{H_1} + \epsilon_{H_2}$. For this, $\epsilon_{Y_1} = (\delta \uv \cdot \delta \uv) \delta u_\ell$ measures the kinetic energy available to be transported by the longitudinal component of $\delta \uv$, $\epsilon_{Y_2} = (\delta \Bv \cdot \delta \Bv) \delta u_\ell$ quantifies the magnetic energy that will be transported by $\delta u_\ell$, $\epsilon_{Y_3} = - 2 (\delta \uv \cdot \delta \Bv) \delta B_\ell$ is related to the velocity-magnetic field correlations coupled to the longitudinal magnetic field fluctuations, $\epsilon_{H_1} = -d_p (\delta \Bv \cdot \delta \Bv) \delta j_\ell /2$ let us know the quantity of magnetic energy advected by longitudinal current-density field fluctuations, and $\epsilon_{H_2} = d_p (\delta \Bv \cdot \delta \jv) \delta B_\ell$ is related to the magnetic-current density field correlations coupled to the longitudinal magnetic field fluctuations. In equation \ref{eq:let}, the two latter terms vanish when $d_p \to 0$, which recovers the classic Yaglom law. 

\begin{figure}
    \centering
    \includegraphics[width=\textwidth]{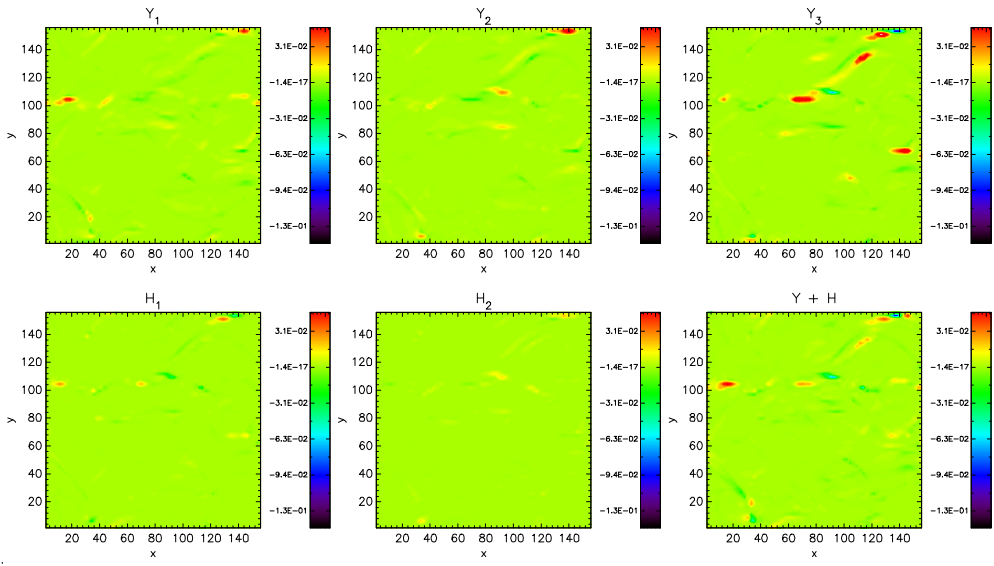}
    \caption{HMHD simulation. Bidimensional maps of Yaglom components: $\Yv_1$ (Eq. \ref{def:y1}), $\Yv_2$ (Eq. \ref{def:y2}), and $\Yv_3$ (Eq. \ref{def:y3}); and Hall components: $\Hv_1$ (Eq. \ref{def:h1}), and $\Hv_2$ (Eq. \ref{def:h2}), of the mean rate of total energy injection, estimated at the scale $\ell\simeq d_p$.}
    \label{fig:yaghall_HMHD}
\end{figure}

\begin{figure}
    \centering
    \includegraphics[width=\textwidth]{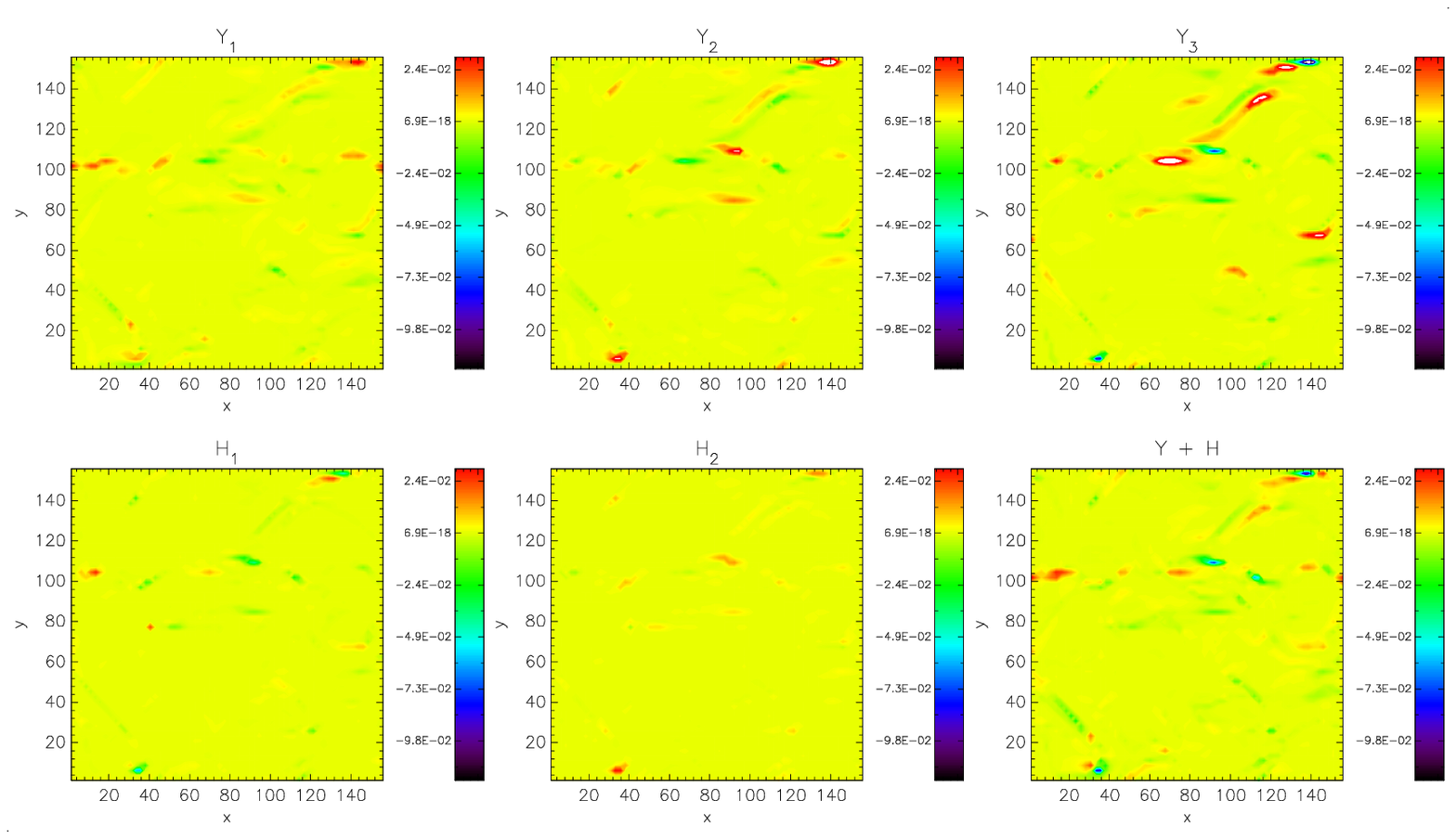}
    \caption{LF simulation. Bidimensional maps of Yaglom components: $\Yv_1$ (Eq. \ref{def:y1}), $\Yv_2$ (Eq. \ref{def:y2}), and $\Yv_3$ (Eq. \ref{def:y3}); and Hall components: $\Hv_1$ (Eq. \ref{def:h1}), and $\Hv_2$ (Eq. \ref{def:h2}), of the mean rate of total energy injection, estimated at the scale $\ell\simeq d_p$.}
    \label{fig:yaghall_LF}
\end{figure}

\begin{figure}
    \centering
    \includegraphics[width=\textwidth]{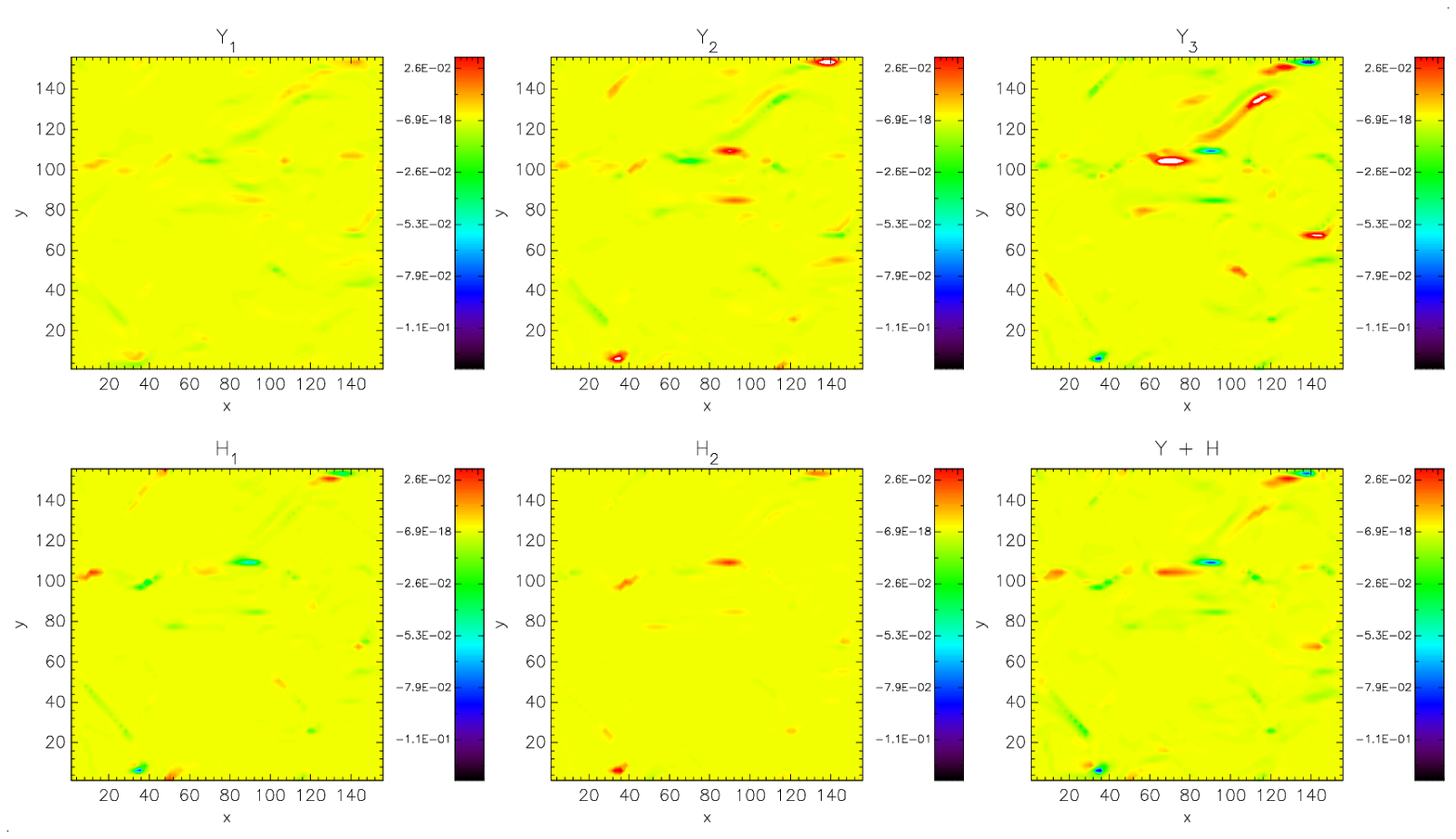}
    \caption{HVM simulation. Bidimensional maps of Yaglom components: $\Yv_1$ (Eq. \ref{def:y1}), $\Yv_2$ (Eq. \ref{def:y2}), and $\Yv_3$ (Eq. \ref{def:y3}); and Hall components: $\Hv_1$ (Eq. \ref{def:h1}), and $\Hv_2$ (Eq. \ref{def:h2}), of the mean rate of total energy injection, estimated at the scale $\ell\simeq d_p$.}
    \label{fig:yaghall_HVM}
\end{figure}

Since we want to explore the energy flux distribution resulting from the turbulent nonlinear transfer, we first focus on the scale $\sim d_p$, {\it i.e.}, at the bottom of the inertial range, where the intermittent structures carrying the most of the energy have been formed. As we noted in Section \ref{dns}, $N_x = N_y = 1024$, then $\Delta x = \Delta y \approx 0.15$, resulting in $\ell \approx 8$. At this scale, we will identify the Yaglom $\Yv_m(x,y) = (\Yv_{mx} + \Yv_{my})/2$, and Hall $\Hv_n(x,y) = (\Hv_{n x} + \Hv_{n y})/2$ contributions, according to their spatial distribution, where $m = 1, 2, 3$ and $n = 1, 2$. These bidimensional maps are presented in Figure \ref{fig:yaghall_HMHD} (HMHD), Figure \ref{fig:yaghall_LF} (LF), and Figure \ref{fig:yaghall_HVM} (HVM). They clearly highlight the intermittent nature of the cross-scale energy transfer. This is evidenced by the presence of strong contributions to the scaling law in small-scale localized, sparse structures. Most of the energy transfer, and consequently of the energy dissipation, is taking place at those locations.
It is interesting to note that different terms of the Hall-MHD law may present similar structures at the same locations. However, in several occasions the structures of different contributions are not co-located, suggesting that the nature of the fluctuations producing the energy transfer may be change with position, possibly enabling different dissipation mechanisms~\citep{sorriso2019turbulence}.
It is also apparent that most of the more intense structures are positive, suggesting that the energy is being transferred towards the small scales in a ``direct'' cascade. However, several negative structures are also present, where the contribution from the various terms is acting as to remove energy from the smaller scales. We do not imply here that the sign of the local contributions is related to the cascade direction, and the latest statements should be taken as a qualitative indication. Note that the overall energy flux will result from the average over the whole domain, and is mostly positive, as expected and as we shall discuss in the next Section.
Comparing the different simulations reveals that the various terms have similar behaviour. In particular, the $\Yv_3$ term seems to be dominating in the simulations. Interestingly, in the HMHD simulation the Hall terms appear weaker than in the two other simulations. This observation will be discussed later.

\begin{figure}
    \centering
    \includegraphics[width=\textwidth]{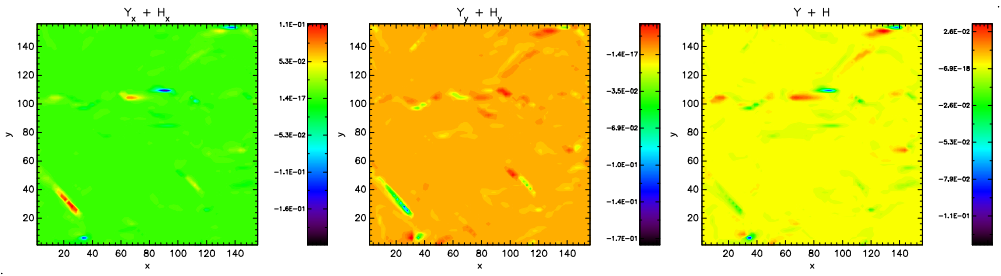}
    \caption{$x$ (left panel) and $y$ (right panel) components of the term $\Yv + \Hv$ (right panel), computed from the HVM simulation, estimated at the scale $\ell\simeq d_p$.}
    \label{fig:anisotropy}
\end{figure}

Finally, the maps of $\Yv + \Hv$ are separated by components in Figure~\ref{fig:anisotropy}, for the HVM simulation (similar behavior was found in HMHD and LF simulations --not shown here). While the general behaviour is similar, some differences between the two components of the energy transfer exist, possibly an effect of the finite size of the ensemble.

\section{Global energy transfer analysis}
\label{global}

\begin{figure}
    \centering
    \includegraphics[width=\textwidth]{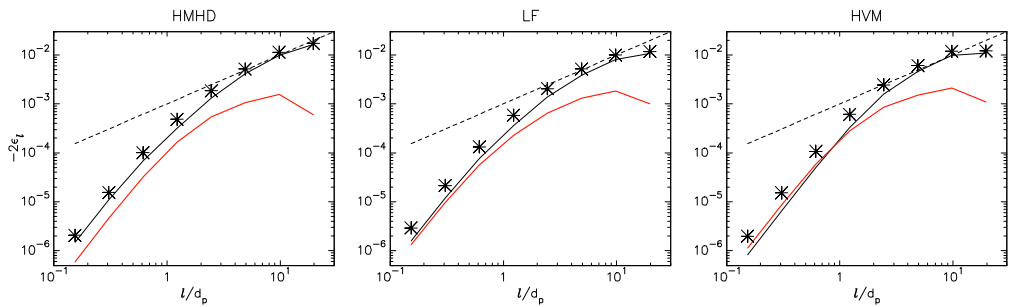}
    \caption{Cascade rate $-2 \epsilon_\ell$ law as a function of the scale $\ell$, conducted in the HMHD (left panel), LF (middle panel), and HVM (right panel) simulations. The contributing terms: $\epsilon_Y$ (black-solid line), and $\epsilon_H$ (red-solid line) are presented together to their direct sum (asterisks). The dashed line represents the linear scaling.}
    \label{fig:yaghall}
\end{figure}

We continue our analysis computing the local turbulent energy transfer rate $- 2 \epsilon_\ell$ across the scales $\ell$. The three panels of Figure \ref{fig:yaghall} present the averaged equation (\ref{eq:let}) over the whole domain, corresponding to the HMHD (left panel), LF (middle panel), and HVM (right panel) simulations. We separate the Yaglom (black-solid line), and Hall (red-solid line) contributions in order to highlight their relative amplitude across the scales $\ell$. The sum of both contributions, giving the Hall-Yaglom law, is plotted with asterisks. As already observed in the previous Section, in the HMHD and LF simulations, the Hall contribution $\epsilon_{H(\ell)}$ is always lower than the Yaglom one. However, both contributions become closer for scales $\ell \lesssim d_p$. In the case of the HVM simulation, that occurs in the range $\ell \lesssim d_p$, $\epsilon_{H(\ell)} > \epsilon_{Y(\ell)}$. Then, at $\ell \approx d_p$, $ \epsilon_{Y(\ell)}$ becomes more relevant and remains so for larger scales. This behavior is consistent with the settings of our simulations, where the mean field is out-of-plane, and there is strong whistler/magnetosonic activity. If we recall that the Hall term $\epsilon_H$ manifests itself through compressible activity, the cascade rates are expected to be affected in the fluid simulations; while in the HVM simulation, this compressible activity is suppressed through in-plane Landau damping and ion-cyclotron resonances, which re-inject in the system incompressible Alfv\'enic-like fluctuations. In this 2D setting, the LF model, in which Landau dissipation is introduced at a quasilinear level, partially quench compressible fluctuations but not as efficiently as the HVM one. In addition, we can confirm that in the interval $2 d_p \lesssim \ell \lesssim 10 d_p$ (roughly corresponding to the MHD-turbulence range), the scaling of $- 2 \epsilon_\ell$ is compatible with a linear scaling law (black-dashed line) for our three cases of study, as previously reported by \citet{sorriso2018local}. 

\begin{figure}
    \centering
    \includegraphics[width=0.9\textwidth]{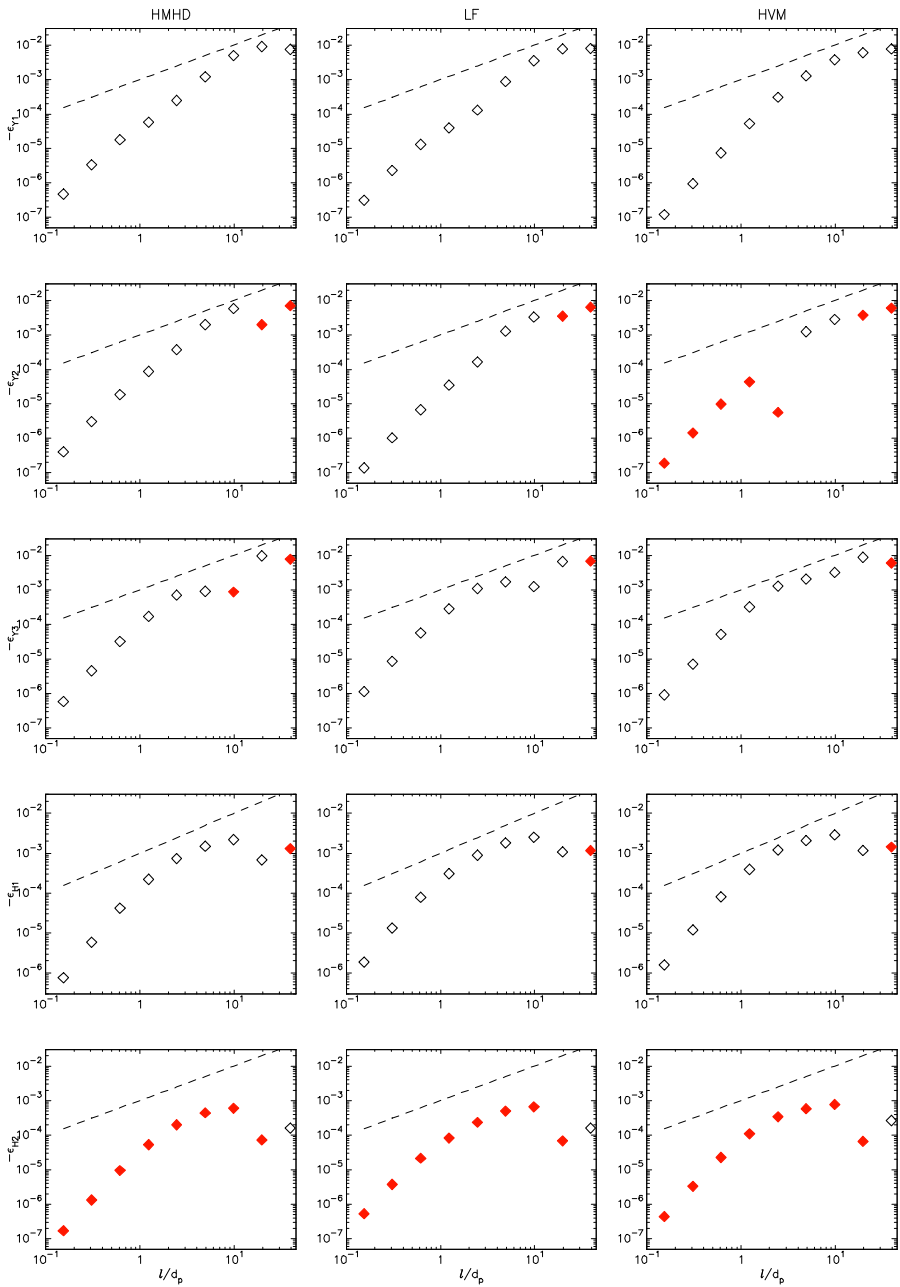}
    \caption{From top to bottom row: $-\epsilon_{Y_1}$, $-\epsilon_{Y_2}$, $-\epsilon_{Y_3}$, $-\epsilon_{H_1}$, and $-\epsilon_{H_2}$. The components are computed from the HMHD (left column), LF (middle column), and HVM (right column) simulations. Black-empty diamonds correspond to the positive sign, while red-filled diamonds show negative values. Linear scaling is plotted with a black-dashed line.}
    \label{fig:epsterms}
\end{figure}

Moreover, in Figure \ref{fig:epsterms} we can see how Yaglom $-\epsilon_{Y_m}$ and Hall $-\epsilon_{H_n}$ terms, respectively, are supporting the latter global behavior. We use black-empty diamonds to picture the positive terms, while negative quantities are plotted with red-filled diamonds. The left column shows HMHD-simulation results, middle column corresponds to the LF simulation, while the results from the HVM simulation are presented in the right column. If we focus on the amplitude and sign of these quantities, we cannot note substantial differences between the individual Hall-contributions among the simulations. However, the sign of $-\epsilon_{Y_2}$ (associated to the magnetic energy transported by the velocity field) is passing from negative to positive at $\sim 2 d_p$ in the HVM simulation. This is not seen in their HMHD and LF counterparts.
%
%

%

\begin{figure}
    \centering
    \includegraphics[width=0.7\textwidth]{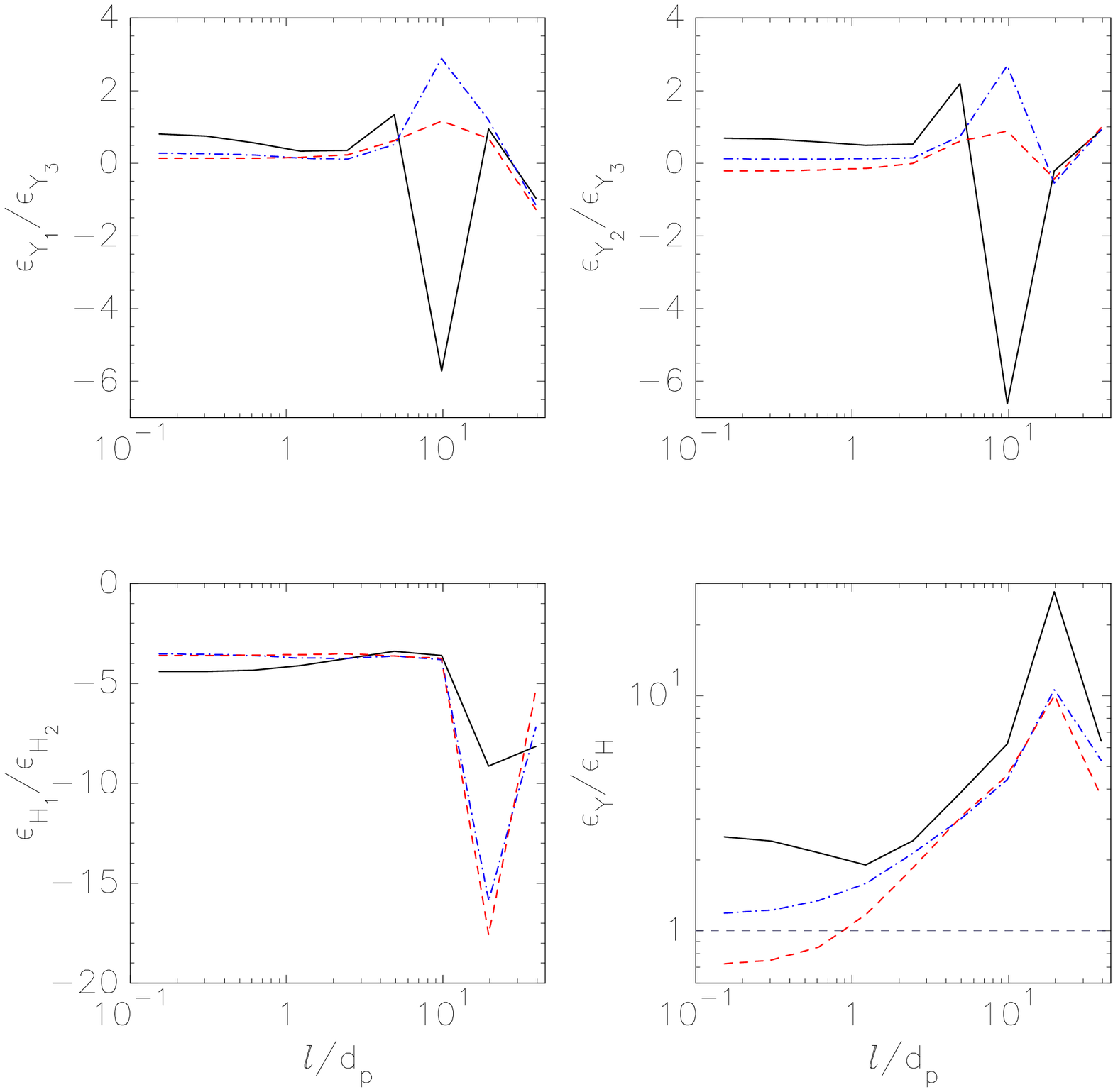}
    \caption{Comparison of the Yaglom terms $\epsilon_{Y_1}$ and $\epsilon_{Y_2}$, respect to $\epsilon_{Y_3}$ (top row). The comparison between the Hall terms ($\epsilon_{H_1}$ and $\epsilon_{H_2}$) is plotted in the bottom-left panel. The bottom-right panel shows the comparison between $\epsilon_{Y}$ and $\epsilon_{H}$. In the latter panel, the horizontal grey-dashed line marks the barrier of $\epsilon_{Y}/\epsilon_{H} = 1$. Black-solid line corresponds to the HMHD simulation, blue-dash-dotted line shows the LF simulation, and HVM simulation is represented by the red-dashed line.}
    \label{fig:compare}
\end{figure}

Finally, we use our results to compare the amplitude of each of Yaglom and Hall terms across the scales. In Figure \ref{fig:compare}, we present these comparisons for the Yaglom terms: $\epsilon_{Y_1} / \epsilon_{Y_3}$ (top-left panel), and $\epsilon_{Y_2} / \epsilon_{Y_3}$ (top-right panel). Hall terms comparison $\epsilon_{H_1} / \epsilon_{H_2}$ can be seen in the bottom-left panel. The bottom-right panel shows $\epsilon_{Y} / \epsilon_{H}$. Black-solid line is for HMHD simulation, blue-dash-dotted line represents the LF results, and the red-dashed line is for the HVM simulation. From a global point of view, we note that the comparisons computed from the LF and HVM simulations are quite similar, specially for scales $\ell \lesssim 10 d_p$. After this range, the sign of $\epsilon_{Y_3}$ fluctuates with respect to that of $\epsilon_{Y_1}$ and $\epsilon_{Y_2}$. On the other hand, the sign of $\epsilon_{H_1}$, as compared with $\epsilon_{H_2}$, remains negative for all of the scales. Once again, LF and HVM simulations seem quite similar when comparing the Hall terms. The comparison $\epsilon_{Y} / \epsilon_{H}$ shows that the HVM simulation is useful to study the transition from fluid to kinetic scales ($\sim d_p$) as we can see when comparing this ratio.       

\section{Discussion and Conclusions}
\label{conclusion}

In this paper we studied the properties of energy transfer contributions in collisionless turbulent plasmas. For this scope, the heuristic proxy LET was computed on three numerical simulations (Hall-MHD, Landau Fluid, and hybrid kinetic Vlasov-Maxwell) of a quasi-steady state of turbulence. Moreover, this study would also be aligned to the ``Turbulent dissipation challenge'' \citep{parashar2015turbulent}, as we test the LET proxy on three different numerical models (which use different numerical schemes) under the same initial conditions, with similar physical and numerical parameters. We have in mind that each of these models have their own characteristic scale. In the HVM model, this scale corresponds to the proton skin depth. In the fluid models this scale corresponds to the dissipative scale, where the energy is dissipated by viscous and resistive effects.\\      
From Figure \ref{fig:yaghall_HMHD} to Figure \ref{fig:yaghall_HVM}, we presented bi-dimensional maps for each of the simulations, and for the scale $\ell = 8$ ($\sim d_p$). In particular, in Figure \ref{fig:anisotropy} we can show 
the location and intensity of the structures where most of the energy is contributing to the cross-scale transfer. 
%
%
\\     
Then, Figure \ref{fig:yaghall} tests the $-2 \epsilon_\ell$ cascade-rate definition (equation (\ref{eq:let})). Here, when only the global contribution of $\epsilon_Y + \epsilon_H$ is plotted (asterisks), no strong differences are seen between our set of simulations. However, when $\epsilon_Y$ (black-solid line) and $\epsilon_H$ (red-solid line) are separated, we note that compressible activity, suppressed (through in-plane Landau damping and ion-cyclotron resonances) in the HVM simulation (right panel), let that $\epsilon_H > \epsilon_Y$ for scales $\ell \lesssim d_p$. This is not reported in fluid-like simulations because $\epsilon_H$ is manifested mainly by the compressible activity. We further point out that a net contribution in the energy budget might be due to the pressure anisotropy, which is not taken into account in the fluid-Yaglom theory.\\
%
We observe that $\epsilon_{Y_2} \sim (\delta \Bv \cdot \delta\uv) \delta u_\ell$ have opposite sign --for the majority of the scales-- in the HVM simulation, with respect to the fluid simulations (Figure \ref{fig:anisotropy}). Also for this parameter, a change of monotony may be noticed in the interval $2 d_p \lesssim \ell \lesssim 10 d_p$ (roughly corresponding to the MHD-turbulence range), only in the HVM simulation. The reason for this behavior is not fully understood.\\
Finally, the comparisons made in Figure \ref{fig:compare} show similitude between our set of simulations for scales $\ell \lesssim 2 d_p$. After this range, the contribution of $\epsilon_{Y_3}$ decreases with respect to  $\epsilon_{Y_1}$ (left-top panel) and $\epsilon_{Y_2}$ (right-top panel). A similar behavior, related to the direction of the energy cascade, is seen when comparing the Hall terms (left-bottom panel), where the amplitude of $\epsilon_{H_2}$ decreases more about one decade respect to $\epsilon_{H_1}$ when $\ell \gtrsim 10 d_p$. \\ 
Our overall conclusion is that the three simulations are similar, but not exactly identical, with respect to the local and global turbulent energy transfer. This implies that a cross-scale interconnection exists between fluid and kinetic dynamics, so that not only the turbulent cascade drives the small-scale kinetic processes, but the latter also control the cascade, acting as a form of dynamical dissipation. Further studies are needed to describe in more details such interconnection.
Work in progress includes the extension of energy-transfer analysis in 3D simulations, and a multifractal study of the turbulence \citep{primavera2019parallel}.\\  

CLV was partially supported by EPN projects: PIM-19-01, PII-DFIS-2019-01 and PII-DFIS-2019-04.


\bibliographystyle{jpp}

\bibliography{main}

\end{document}